\documentclass[reprint, amsmath,amssymb,aps,floatfix]{revtex4-2}
\usepackage{hyperref}
\usepackage{graphicx}
\usepackage[usenames,dvipsnames,svgnames,table]{xcolor}
\usepackage{soul} %for highlighting stuff
\usepackage{gensymb}

\begin{document}

\title{Strain coupling of a single exciton to a nano-optomechanical resonator}

\author{Matteo Lodde}
\affiliation{Department of Applied Physics and Science Education, Eindhoven Hendrik Casimir Institute, Eindhoven University of Technology, 5612 AZ, Eindhoven, Netherlands}
\author{Ren\'e P. J. van Veldhoven}
\affiliation{Department of Applied Physics and Science Education, Eindhoven Hendrik Casimir Institute, Eindhoven University of Technology, 5612 AZ, Eindhoven, Netherlands}
\author{Ewold Verhagen}
\affiliation{Center for Nanophotonics, AMOLF, Science Park 104, 1098 XG Amsterdam, The Netherlands}
\affiliation{Department of Applied Physics and Science Education, Eindhoven Hendrik Casimir Institute, Eindhoven University of Technology, 5612 AZ, Eindhoven, Netherlands}
\author{Andrea Fiore}
\affiliation{Department of Applied Physics and Science Education, Eindhoven Hendrik Casimir Institute, Eindhoven University of Technology, 5612 AZ, Eindhoven, Netherlands}

\begin{abstract} 
We demonstrate coupling of a semiconductor quantum dot (QD) to an optomechanical cavity, mediated by the strain of a nano-mechanical mode. The device comprises an optomechanical photonic crystal nanobeam in GaAs with embedded In(Ga)As QDs. The flexural mechanical mode of the device can be optically driven exploiting the large optomechanical coupling rate of the cavity. The vibrations generate a time-modulated strain field that shifts the quantum dot transition energy. We observe that optical driving of the mechanical mode induces a shift in an excitonic line corresponding to an estimated vacuum strain coupling rate of 214 kHz. Our approach represents an important step towards the use of phonons to couple different on-chip quantum systems.
\end{abstract}

\maketitle

\section{Introduction}
Recent developments in electro- and optomechanics have proven that phonons in engineered nanostructures offer versatile capabilities for precision sensing and quantum information processing \cite{Treutlein2014}. Indeed, they can coherently couple to different quantum systems, including spin qubits\cite{Ovartchaiyapong2014, Khanaliloo2016, meesala_enhanced_2016, wang_coupling_2020}, excitons in quantum dots (QDs)\cite{Gell2008, Yeo2014, Montinaro2014, weis_surface_2016, sollner_deterministic_2016, yuan_frequency-tunable_2019, wigger_resonance-fluorescence_2021, choquer_quantum_2022, Spinnler2024}, optical cavities\cite{eichenfield_picogram-_2009, Chan2009, Chan2011, aspelmeyer_cavity_2014, gomis-bresco_one-dimensional_2014, leijssen_strong_2015} and ions \cite{jean-franccedil_microscale_2019}, making them good candidates for transduction of quantum information between disparate degrees of freedom.
Nanostructures that host highly coherent mechanical excitations include on the one hand localized mechanical resonators, which can feature extremely low dissipation \cite{Engelsen2024}, and on the other hand phononic waveguides that feature low propagation losses, with acoustic waves e.g. confined by phononic crystals on a chip \cite{Patel2018, Zivari2022}.

Coupling an excitonic state in a single QD –- which at low temperatures can be considered as a two-level system (TLS) -– to a mechanical resonator establishes an interaction in which the intrinsic non-linearity of the exciton in principle allows creating and manipulating non-classical states of the mechanical oscillator. This is particularly interesting for phononic quantum technologies \cite{Barzanjeh2021, Carter2017} in which the creation and manipulation of single phonons would play an enabling role. Recently, several hybrid systems involving a mechanical resonator and a QD have been explored, exploiting different materials and architectures such as micropillars, cantilevers, nanobeams and membranes \cite{Yeo2014, Montinaro2014, weis_surface_2016, Munsch2017, Kettler2020, Vogele2020, wigger_resonance-fluorescence_2021}. Some of these works focus on measuring Brownian motion \cite{Spinnler2024, Munsch2017}, other studies effectively enhance the coupling strength by electrically driving the mechanical resonator through external piezos or interdigital transducers \cite{Gell2008, Yeo2014, Montinaro2014, weis_surface_2016, Vogele2020, wigger_resonance-fluorescence_2021}, and a third class induces the mechanical motion by resonantly exciting the QD \cite{Kettler2020}.

Here we demonstrate a system in which both the drive and the readout are purely optical. To achieve that, our device combines three different systems: an optical cavity, a mechanical resonator and a TLS. As such, it forms a tripartite system with interactions as illustrated in Fig. \ref{fig:1}A, where the mechanical resonator acts as intermediate bus between the optical cavity and the TLS. By resonantly driving the mechanical mode with a modulated laser field we demonstrated a strain-driven shift of the excitonic emission energy up to $4.9$~GHz, corresponding to a vacuum strain coupling rate of $214$~kHz. The employed optical driving does not require electrical contacts to be defined on or close to the chip and may provide additional design flexibility in experiments involving phonons and TLS. Moreover, the realized interaction represents a useful addition to the growing toolbox for the coherent control of phonons. 

\section{Methods and materials}
\subsection{Device concept}
\begin{figure}
    \centering\includegraphics[width=1\linewidth]{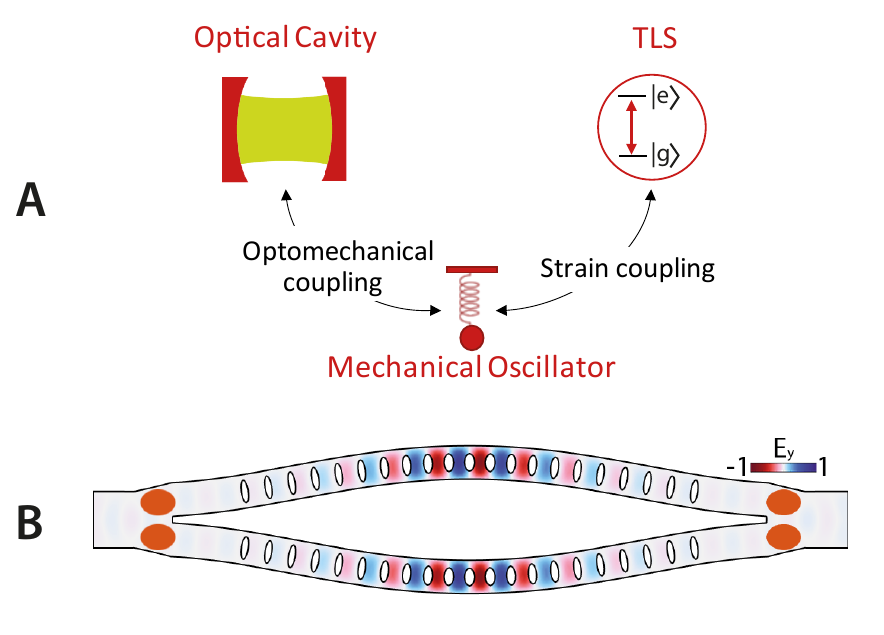}
    \caption{(\textbf{A}) Sketch of the systems and their interactions exploited in the device. The optical cavity is coupled to the mechanical resonator via a parametric radiation pressure coupling, while the mechanical oscillator couples to the TLS via the mechanical strain field. (\textbf{B}) Implementation of the interactions of (A) in a nanodevice constituted by two coupled nanobeams. The color scheme represents the fundamental symmetric optical mode, the deformation (highly enhanced for visibility) represents the mechanical vibration of the two nanobeams, and the orange circles indicate the ideal position for the QDs, corresponding to the areas where the strain is maximal.}
    \label{fig:1}
\end{figure}

The proposed implementation of the tripartite system and its interactions is shown in Fig. \ref{fig:1}B. 

This device is based on a zipper cavity \cite{Chan2009}, an optomechanical cavity consisting of two double-clamped nanobeams with a photonic crystal cavity (PhCC) patterned them. The two cavities are coupled through their evanescent fields, resulting in hybrid modes whose frequency depends on the beam separation. This is associated with a gradient force which couples to the flexural mechanical motion of the beams. The coupling between the mechanical resonator and the QDs is achieved via the strain induced by the vibrations of the nanobeams, which shifts the energy levels of the QDs proportionally to the local strain \cite{chuang2012physics}. The ideal position of QDs to observe such coupling is close to the clamping points, as there the strain is maximum. Moreover, the optical field in the clamping points is almost negligible, preventing a direct interaction between the QD and the optical field, via the Jaynes-Cummings Hamiltonian or through the creation of free carriers. The read out of the QD is achieved by means of micro-photoluminescence measurements in which a second laser -- not resonant with the optical cavity -- is used.

\subsection{Coupling between the systems}
The device thus exploits two coupling mechanisms: the optomechanical coupling between the optical cavity and the mechanical resonator, and the strain coupling between the mechanical resonator and the QD.
The strength of the optomechanical interaction is quantified as the shift of the optical frequency ($\omega_\mathrm{cav}$) per unit of the displacement ($x$), normalized by the amplitude of the zero-point fluctuations ($x_\mathrm{zpf}$) of the resonator:
\begin{equation}
    g_0 = -\frac{\partial \omega_\mathrm{cav}}{\partial x} x_\mathrm{zpf}
\end{equation}
This parametric coupling allows resonant driving of the mechanical mode via radiation pressure by means of an intensity-modulated laser field. A full intensity modulation of the laser resonant with the mechanical mode produces, in the linear regime, a number of phonons $n_\mathrm{p}$ given by
\begin{equation}
    n_\mathrm{p} = \frac{2g_0^2 n^2_\mathrm{cav}}{\Gamma_\mathrm{m}^2}
\end{equation}
with $\Gamma_\mathrm{m}$ being the mechanical energy decay rate, and $n_\mathrm{cav}$ is the photon number in the optical cavity. The full derivation of this equation is available in the Supplementary Material.

The coupling between the QD and the mechanical mode is mediated by strain. The flexural vibrations of the structure induce a time-modulated strain field that modulates the QD transition energy.
The strain-induced shift of QD excitonic transition energies is approximately given by the shift of the bulk transition energy (difference between the conduction and heavy-hole band edges), equal to \cite{chuang2012physics}:
\begin{equation}
    \Delta E = a(S_{xx} + S_{yy} + S_{zz}) + \frac{b}{2}\left(2 S_{zz} - S_{xx} - S_{yy} \right)
\end{equation}
where $a=-8.33$~eV and $b=-2$~eV are the deformation potential constants, related to the isotropic and shear strain in GaAs\cite{Vurgaftman2001} respectively, $S_{ii}$ denotes the strain component along the $i$ direction, and $z$ is the quantization axis (the growth direction in our case). The strain field can be expressed in terms of quantized mechanical eigenmodes. For a single mechanical mode, the strain interaction between a phonon and a TLS is given by the following spin-boson Hamiltonian \cite{Yeo2014}:
\begin{equation}
    \hat{H}_\mathrm{strain} = \hbar g_\mathrm{QD} \hat{\sigma}_z \left(\hat{b} + \hat{b}^\dag \right)
\end{equation}
with $\hat{\sigma}_z$ being the Pauli operator related to the TLS, $\hat{b}$ the phonon annihilation operator and $g_\mathrm{QD}$ the vacuum strain coupling rate between the QD and the mechanical mode:
\begin{equation}
    g_\mathrm{QD} = \frac{1}{\hbar} \left. \frac{\partial E}{\partial x} \right \rvert_{x=0} x_\mathrm{zpf}
\end{equation}
The value of $g_\mathrm{QD}$ can be numerically evaluated by simulating the strain profile of the structure for a normalized displacement, as shown in Fig. \ref{fig:2}D, scaling the mode such that the total mechanical energy is set to the vacuum energy $\hbar\omega_\mathrm{m}/2$, where $\omega_\mathrm{m}$ is the mechanical frequency.
Due to the normalization to the vacuum energy, the total energy shift of the emission of the exciton when the phonon population in the resonator is $n_\mathrm{p}$ is given by \cite{Leijssen2017}:
\begin{equation}
    \Delta E =\hbar g_\mathrm{QD}\sqrt{2n_\mathrm{p}}
\end{equation}
where $n_\mathrm{p}$ is the number of phonons in the resonator. 

\subsection{Simulation and design}
\begin{figure}
    \centering\includegraphics[width=1\linewidth]{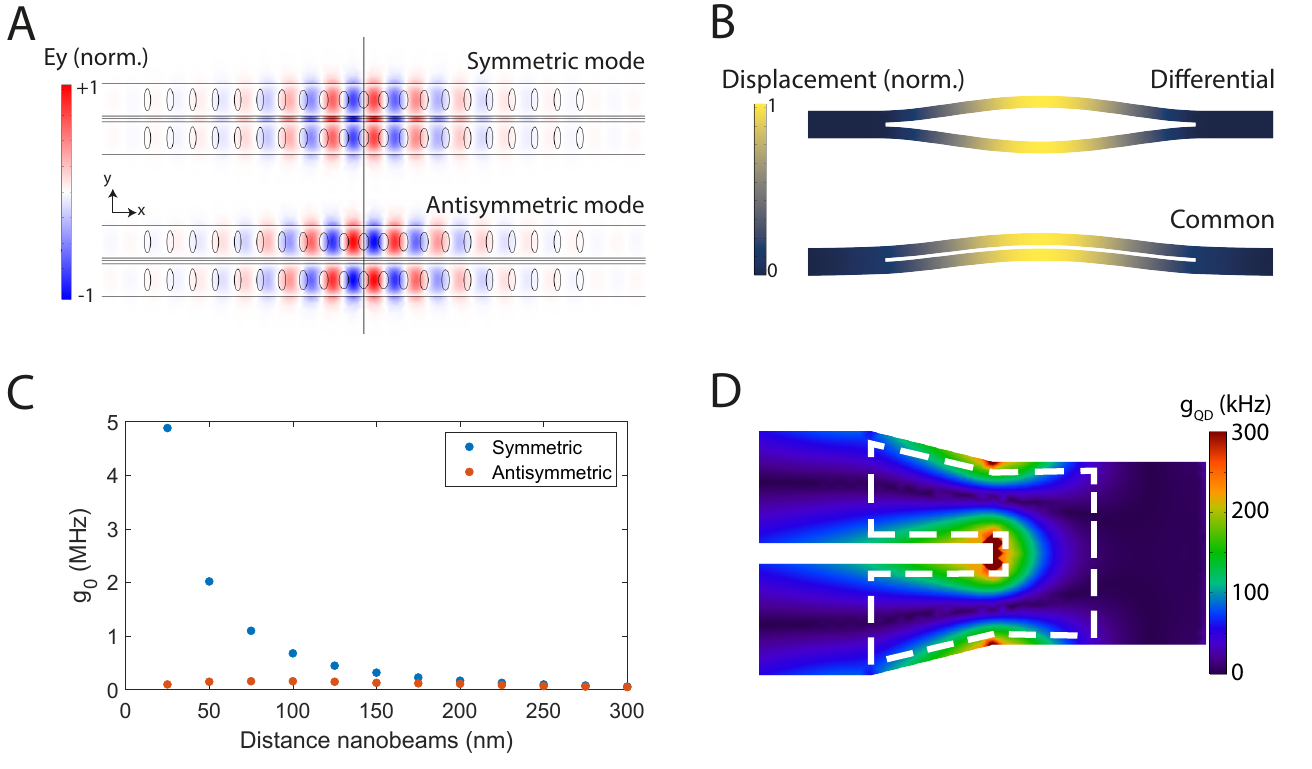}
    \caption{\textbf{A} Optical simulation of the symmetric and antisymmetric modes. \textbf{B} Mechanical simulation of the flexural modes of double-clamped nanobeam, showing the differential mode (top) and the common mode (bottom). The color scale represents the normalized displacement. The amplitude of the motion has been greatly exaggerated for clarity and the holes of the PhCC have not been considered in the simulation. \textbf{C} Simulation of the optomechanical coupling between the symmetric (antisymmetric) optical mode and the differential mechanical mode in blue (red). \textbf{D} Simulation of the strain coupling rate in the clamping point of the nanobeam for the differential mechanical mode. The white dotted lines define the region where the strain has been optimized.}
    \label{fig:2}
\end{figure}

The device consists of a zipper photonic crystal cavity (PhCC) \cite{Chan2009} made in a 250 nm-thick GaAs membrane, which is realized by the coupling of two one-dimensional PhCCs defined in two identical doubly-clamped nanobeams. Each one-dimensional PhCC is realized by patterning elliptical holes in the 530 nm-wide nanobeam. The optical cavity is formed by a gradual change of the hole parameters (size, ellipticity and lattice constant reported in Supplementary) between the outer crystal cell and the defect in the center \cite{Chan2012}, resulting in a localized optical mode. By placing the two nanobeams within the near field of each other the optical modes of each PhCC hybridize, splitting into symmetric and antisymmetric supermodes. A finite element method (FEM) simulation (COMSOL Multiphysics) of the electric field distribution of these two modes of the coupled system is shown in Fig. \ref{fig:2}A. 
The linewidth of the optical mode sets a maximum value for the number of phonons that can be pumped. This can be estimated by the criterion that the shift of the optical cavity due to the dispersive optomechanical coupling needs to be smaller than its linewidth $\kappa$, which amounts to $n_\mathrm{p} < (\kappa / g_0) ^2$. This brings to the counter-intuitive result that a lower optical $Q$ (provided enough laser power) results into a higher achievable maximum phonon population. For this reason the number of photonic crystal mirrors has been chosen to keep the $Q$ factor below $10^4$. Specifically, the simulated values are $Q_\mathrm{symm} = 6425$ and $Q_\mathrm{antisymm} = 9144$.

The device is side-coupled to a free-standing waveguide (in blue in Fig. \ref{fig:3}C). This allows coupling light into the device and measuring the light reflected by it using a lensed fiber. To minimize the back-reflections at the air interface, the tip of the waveguide has been reduced to 200 nm, resulting in an effective mode index close to 1 \cite{Kirsanske2017}. Then, the width is adiabatically increased to match that of the nanobeam and terminated with a photonic crystal mirror tuned to the PhCC’s resonance wavelength. The first 8 holes of the mirror are tapered (their dimensions are indicated in the Supplementary Materials). The offset between the center of the cavity and the first hole of the mirror corresponds to 328~nm and has been chosen to maximize the coupling rate. The coupling rate between the waveguide and the zipper cavity can be tuned by changing their distance. We choose to operate in the critical coupling regime, to increase the visibility of the reflected signal from the cavity.

Also the two fundamental mechanical modes can be seen as two supermodes, which we refer to as common, when the two nanobeams are moving in phase, and differential, when the two nanobeams are moving in opposite phase, as depicted in Fig. \ref{fig:2}B. The simulation has been carried out with a simplified geometry, where the holes have been removed as they have a minor influence on the mechanical displacement profile. Following the idea of the “notched” design in Ref. \cite{gala_nanomechanical_2022}, the frequency between the common and differential mode are split by making the slot between the two nanobeams shorter than the total length. In this way, the common mode is pushed to lower frequency (because the entire length of the suspended beams is involved in the motion) while the frequency of the differential mechanical mode is unperturbed.

As in the common mode the two nanobeams move together, their distance stays constant throughout the whole oscillation and the effect on the optical mode is expected to be small. Vice versa, in the differential mode the slot between the two beams changes, with a large effect on the frequency of the optical mode. By following the steps of \cite{Chan2012} we numerically evaluated the optomechanical coupling. The simulated optomechanical coupling between the differential mechanical mode and both the symmetric and antisymmetric optical modes is reported in Fig. \ref{fig:2}C. As expected from the distribution of the electric field, the highest coupling between the optical mode and the mechanical mode is achieved with the symmetric optical mode. This coupling rate decreases exponentially with the distance between the two nanobeams.

We now consider the strain coupling of QDs to the differential mechanical mode, which can be efficiently driven optically. In a flexural deformation of a symmetric double-clamped beam, the bending produces strain of opposite signs along two opposite surfaces, with a continuous variation between them. This implies that the line passing through the center of the beam, called neutral axis, does not undergo any longitudinal strain, while the strain profile is maximum at the edges. For this reason, our focus has been directed towards optimizing the clamping points of the nanobeam, with the aim of maximizing the strain in a relatively large area. This can be achieved by tapering off the clamping points, as shown in Fig. \ref{fig:2}D. Clearly, as the tether holding the beam becomes thinner, the strain increases and gets localized. Apart from the constraints imposed by fabrication -– which establish a minimum width for the supporting tether –- also the density of the QDs used in the study determines how localized the strain profile should be. Indeed, the higher the strain localization, the smaller is the probability of finding a QD in a region with high strain. Another constraint is imposed by the fact that the photoluminescence (PL) of QDs which are close to the surface is degraded due to non-radiative channels from the presence of surface states \cite{Miao2005}. For this reason, we limit our attention to what we define the \textit{optically active} area of the nanobeam (denoted as $A$ in Eq. \ref{eq:FOM}  below), i.e. the area of the clamping point $\sim$50 nm far from the edges, indicated by the white dotted line in the sketch on Fig. \ref{fig:2}D.

To enhance the strain in this area we taper the ends of the nanobeam, with the goal of optimizing the average strain coupling. The parameters that are optimized are the length and width of the taper at the end of the nanobeam, and the figure of merit (FOM) that is maximized by the algorithm is the surface average of the absolute value of vacuum strain coupling over the area $A$:

\begin{equation}\label{eq:FOM}
    \mathrm{FOM} = \frac{1}{A} \iint_A |g_\mathrm{QD}| \  dx \  dy
\end{equation}
The result of the optimization as well as the absolute value of the vacuum strain coupling is shown in Fig. \ref{fig:1}D. The simulation shows that within the \textit{optically active} area we can achieve a vacuum strain coupling rate up to 250~kHz. Given the high localization of the strain profile, only few QDs are expected to show a measurable strain coupling. Indeed, the surface for which $g_\mathrm{QD}>150$~kHz is 0.069 $\mu\mathrm{m}^2$ which is expected to contain on average $\sim 1$ QD (considering an areal density of 15~dots/$\mu$m$^2$ used for this experiment).
This value could be even lower, as we do not precisely know the minimum QD-surface distance for which our QDs display narrow emission lines. Indeed,  InGaAs QDs have been reported to be sensitive to charge fluctuations of defects located within $\sim$100~nm \cite{houel_probing_2012}.

The simulated value for the vacuum strain coupling rate is comparable to other reports of strain coupling in different mechanical structures, with values in the range between 68~kHz to 3.6~MHz \cite{Yeo2014, Montinaro2014, Munsch2017, Kettler2020, choquer_quantum_2022, Spinnler2024}.
Higher $g_\mathrm{QD}$ values could be obtained with breathing mode mechanical resonances in the GHz range (see Supplementary Materials), but that the proposed structure allows a simpler separation of the optical and mechanical functionalities.

\subsection{Growth and fabrication}
\begin{figure}[ht!]
    \centering\includegraphics[width=1\linewidth]{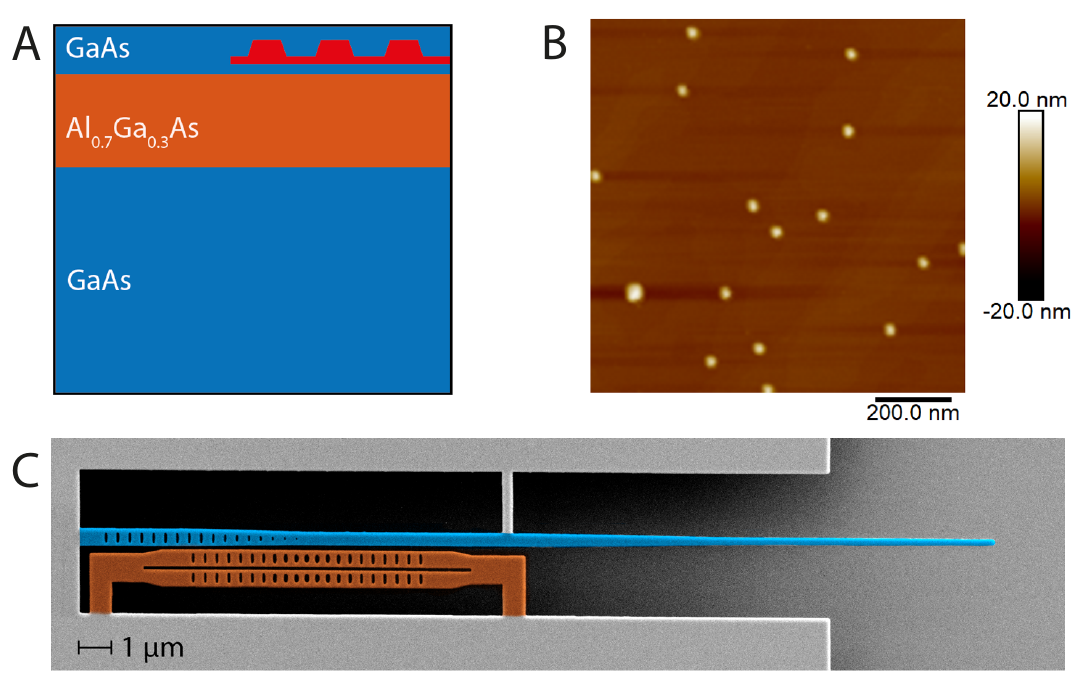}
    \caption{\textbf{A} Wafer layer stack employed to fabricate the devices. The low-density QDs are represented in red. \textbf{B} AFM image of a reference wafer with uncapped dots used to measure the surface density. \textbf{C} SEM image of the fabricated device. In blue: Waveguide used to couple the light into the device, in red: zipper cavity.}
    \label{fig:3}
\end{figure}

The wafer employed in this work consists of a 250 nm-thick GaAs membrane on top of a 3 $\mu$m-thick Al$_{0.7}$ Ga$_{0.3}$As sacrificial layer grown on a nominally undoped (001) GaAs wafer. In the middle of the membrane we integrated a single layer of self-assembled InAs QDs. The structure of the layer stack is shown in Fig. \ref{fig:3}A. The growth has been carried out by molecular-beam epitaxy and the growth parameters have been tuned to ensure a low surface density of approximately 15~dots/$\mu$m$^2$ and a room (low) temperature emission centered around 1280 nm (1220 nm), following the steps of Ref. \cite{Alloing2005}. The surface density has been confirmed with atomic force microscope (AFM) of uncapped QDs samples (Fig. \ref{fig:3}B).

Prior to the fabrication, the wafer was cleaned with isopropanol and acetone and de-oxidized by using a solution of HCl:H$_2$O in a 1:1 ratio for 4 minutes. A hard mask of silicon nitride (250~nm) was deposited using plasma-enhanced chemical vapor deposition. The pattern was defined by electron beam lithography in a 400~nm-thick ZEP 520A resist, and subsequently transferred first to the silicon nitride mask and then to the GaAs membrane using reactive-ion etching. The membrane was then suspended by selectively etching the sacrificial layer using a solution of 10\% HF in water at room temperature.

\subsection{Experimental setup}
\begin{figure}[ht!]
    \centering\includegraphics[width=1\linewidth]{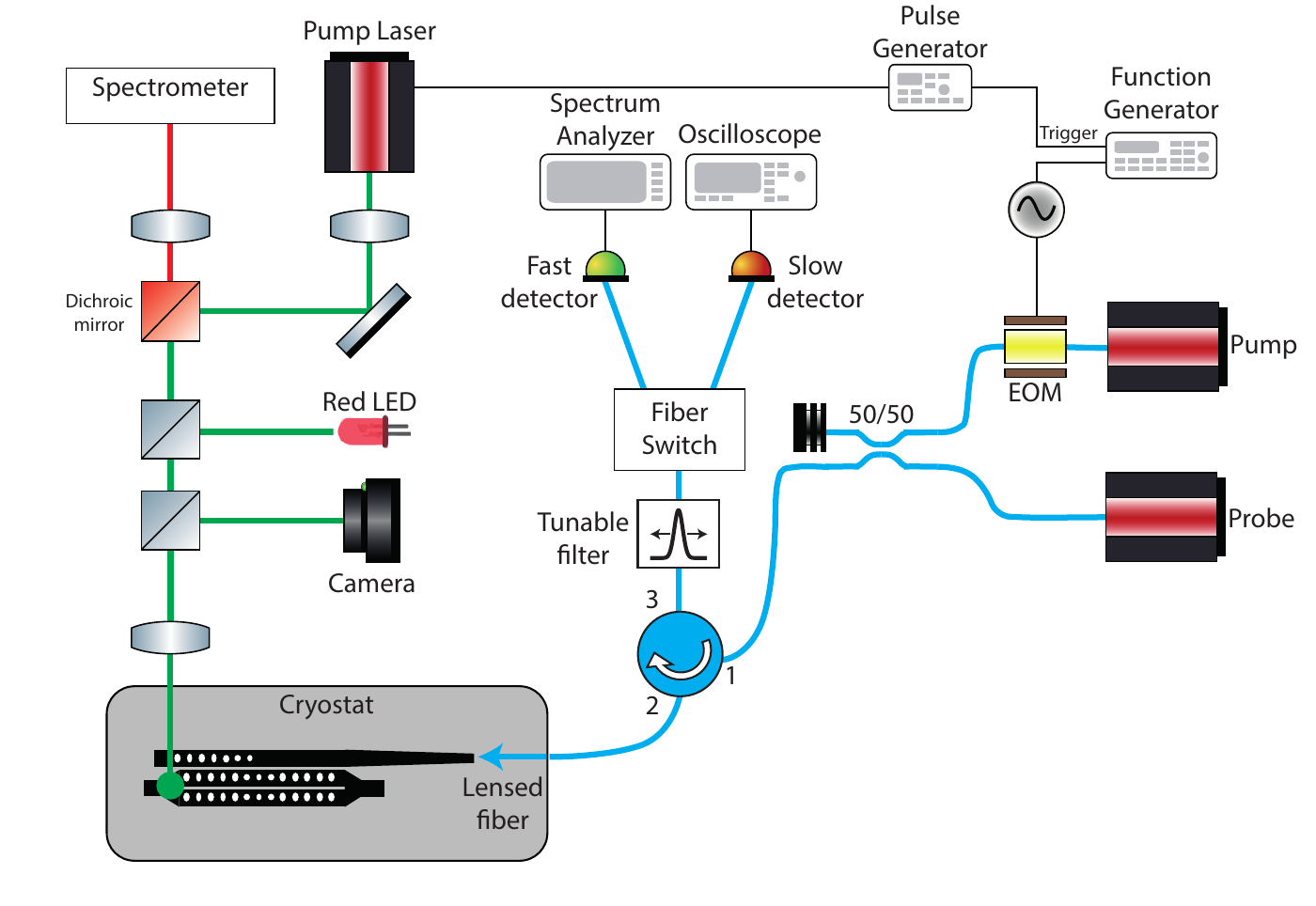}
    \caption{Schematic representation of the experimental setup. Blue lines represent optical fibers, while red and green lines represent the free-space optical path. The solid black lines represent electrical connections.}
    \label{fig:4}
\end{figure}

The experimental setup employed in this work is schematically represented in Fig. \ref{fig:4}. It allows carrying out micro-photoluminescence (micro-PL) measurements from the top while simultaneously coupling light into the cavity via a lensed fiber from the side. The sample is placed in a continuous flow cryostat which can be cooled down to approximately 10 K using liquid helium.

The cavity reflection measurements from the side are carried out using two tunable lasers with emission in the range between 1510 nm to 1610 nm. One is used as a probe (tuned to the antisymmetric optical mode to monitor the mechanical displacement) while the other is used as a pump (tuned to the symmetric optical mode) to drive the mechanical oscillation. The pump laser is amplitude-modulated using an electro-optic modulator (EOM) which is driven by a function generator. Both lasers are fiber coupled and combined using a 50/50 fiber coupler. The laser light is then sent to the lensed fiber via a circulator. Since the waveguide supports only a transverse-electric (TE) mode, to ensure a good coupling, all the fibers and fiber-coupled devices up to the lensed fiber are polarization-maintaining. The detection of the reflected probe power, after suppressing the pump with a tunable filter, can be achieved by monitoring the power in the third circulator port either by using a slow detector read by an oscilloscope or a fast detector (bandwidth of 250~MHz) connected to a spectrum analyzer. 

The micro-PL part of the setup consists of a cage system held above the cryostat window and controlled by a x-y-z piezo stage. This allows aligning the optical setup to the desired position of the sample while keeping the alignment between the lensed fiber and the waveguide fixed. The sample is excited with a pump laser (980 nm) which can either be used in continuous-wave operation to characterize the PL of the sample, or in a pulsed mode to perform stroboscopic measurements. In the latter case, the laser diode is connected to a pulse generator (with a pulse duration set to 7.15 ns) that is triggered by the function generator, thus synchronizing the mechanical drive with the PL measurement. The laser light is focused onto the sample using an objective lens (Mitutoyo, with numerical aperture NA=0.42) providing a spot size of about 8 $\mu$m$^2$. The PL signal is collected via the same objective and split from the pump path using a dichroic mirror with a cutoff wavelength at 1000 nm. The PL signal is finally coupled to a single-mode fiber connected to a spectrometer (with a Gaussian instrument response function characterized by a $\sigma_\mathrm{PL}=4$ GHz). Finally, the setup includes a red LED as a broad light source to illuminate the sample and a CCD camera to image the sample for the alignment.

\section{Results}
\subsection{Optomechanical characterization}
\begin{figure}[ht!]
    \centering\includegraphics[width=1\linewidth]{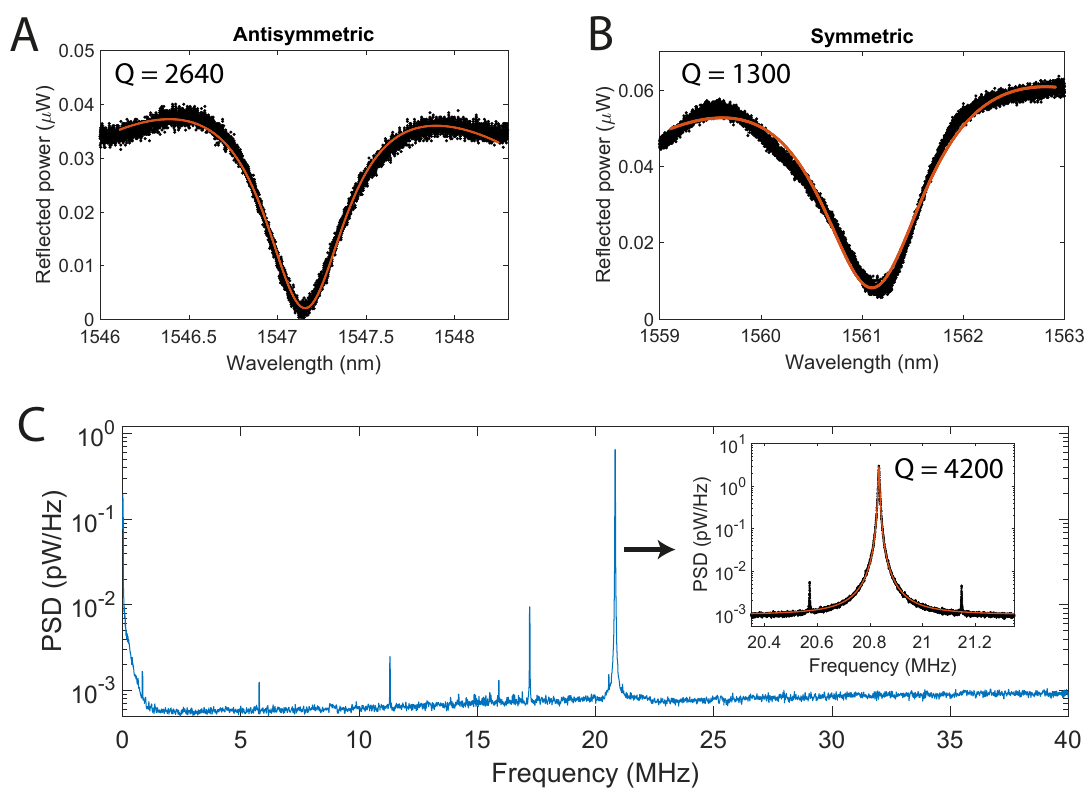}
    \caption{Low temperature measurement and Lorentzian fit of the antisymmetric (\textbf{A}) and symmetric (\textbf{B}) optical resonance. (\textbf{C}) Low temperature measurement of the mechanical power spectral density (PSD) using the symmetric optical mode. The inset shows a detailed measurement and Lorentzian fit of the differential mechanical mode.}
    \label{fig:5}
\end{figure}

The devices were first characterized at low temperature, to measure their optical and mechanical resonances. The optical cavity was measured in reflection by sweeping the wavelength of the probe laser, with the pump laser off, and measuring the reflected signal using the slow detector. In the laser range, we can clearly identify two optical resonances, corresponding to the antisymmetric and symmetric optical supermodes (Fig. \ref{fig:5}A-B). 
The experimental spectra were fitted using a Lorentzian fit with a quadratic polynomial background, from which we can extract the total loss rate of the cavity $\kappa$. For these measurements the probe laser was set at its minimum power (corresponding to $\sim 0.5~\mu$W injected into the waveguide) to minimize the optical bistability effect. The amplitude of the reflection dip, that drops to almost zero in resonance, confirms that the cavity is approximately critically coupled, as stated by the expression for the minimum cavity reflectance $R_\mathrm{min}=(\kappa_0 - \kappa_\mathrm{ex})^2/\kappa^2$, $\kappa_0$ being the intrinsic loss rate and $\kappa_\mathrm{ex}$ being the extrinsic loss rate due to the coupling to the waveguide, and where $\kappa = \kappa_0 + \kappa_\mathrm{ex}$.
The intrinsic losses are influenced by fabrication imperfections, which reduce the experimental $Q$ factor by a factor 4 compared to the simulation (values reported in Fig. \ref{fig:5}A and B).

By tuning the laser to the side of the optical resonance, the mechanical vibrations which shift the optical spectrum are transduced into an intensity modulation, that is measured with the fast detector and analyzed with the spectrum analyzer. Fig. \ref{fig:5}C reports the mechanical spectrum obtained by setting the laser to the sideband of the symmetric optical mode. From the simulation of the device we expect the common (differential) mode to have an eigenfrequency of 15.3 MHz (17.8 MHz). By comparing the spectrum with the simulated mechanical eigenmodes we can clearly identify the common and differential mechanical mode (the one with resonant frequency $\omega_\mathrm{m}=20.85$ MHz) as the two main peaks, where the differential mode shows a much larger transduction, as expected.
We suspect the discrepancy between the simulated and experimental frequency to be due to fabrication imperfections and to the temperature dependence of the Young modulus. In the inspected range we also observe other peaks, whose amplitude is 3 orders of magnitude smaller than the main peak. These are related to the presence of other flexural modes in the structure, which are weakly coupled to the optical mode. Due to the absence of a phononic shield -- not feasible for this frequency range, due to the large required dimension of the phononic crystal -- the losses (for the differential mode $Q_\mathrm{m}=4200$) of the mechanical modes are limited by the clamping losses, i.e. acoustic radiation into the bulk.

The vacuum optomechanical coupling rate has been estimated by measuring the driven phonon population at different laser power (see Supplementary Materials), providing an experimental value of $g_0/2\pi = 363\pm25$~kHz, while the simulated value corresponds to 427~kHz.

\subsection{Stroboscopic measurements of strain coupling}
After characterizing the cavity, the strain coupling strength has been measured by means of stroboscopic PL measurements. The IR pump laser has been set to the symmetric optical mode and modulated at the mechanical frequency using the EOM. By performing a pump-probe measurement we can ensure that the IR pump laser is driving the mechanical resonator at its resonant frequency $\omega_m$. The average power of drive laser coupled to the lensed fiber is 1.5~mW.
From the coupling efficiency between the lensed fiber and the waveguide, and the extrinsic loss rate of the cavity, we estimate the photon number in the symmetric optical mode as being $n_\mathrm{cav}\approx2500$.

The IR optical drive is synchronized with the laser pump used to perform the stroboscopic PL measurement. The PL laser pump emits pulses with a duration (FWHM) of 7.15 ns, corresponding to a duty cycle of 15\% for a modulation frequency of 20.85 MHz. This pulse duration has been chosen as a trade-off between the PL counts and the time resolution needed to observe the synchronous shift of the exciton energy. The average power of the PL pump incident on the PhCC corresponds to 6~$\mu$W. The PL measurement is conducted in a region of the sample close to the clamping point of the nanobeam, exploiting the optimized region of the device. The relative phase between the modulation of the driving laser and the optical pump was swept between 0 and 2$\pi$ with a step of $\pi/4$, while the modulation frequency of both was slightly readjusted in each measurement to match the frequency of the thermal peak and compensate for slow drift of the mechanical frequency. The spectra are shown in Fig. \ref{fig:6}A, in which 5 peaks are highlighted.

\begin{figure}
    \centering\includegraphics[width=1\linewidth]{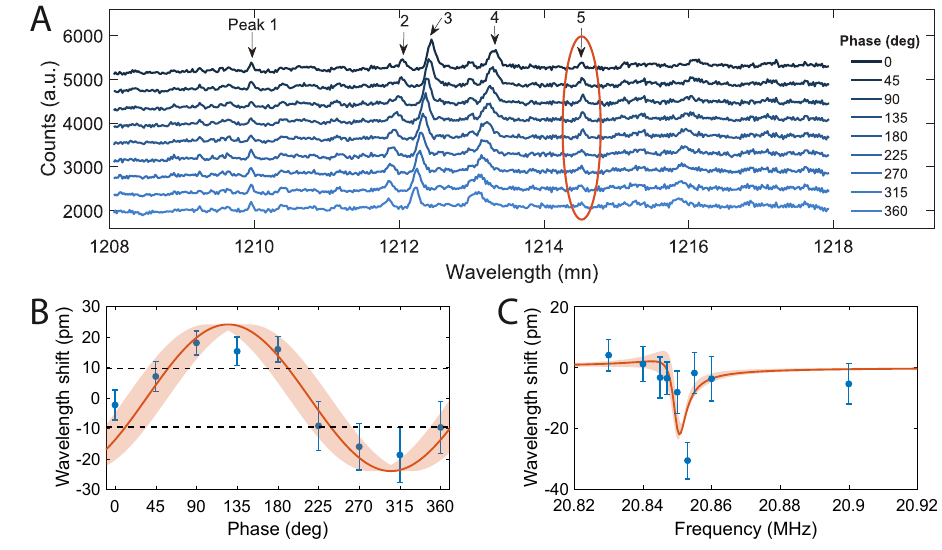}
    \caption{\textbf{A} PL measurements of the QDs in the clamping point of the device taken at different driving phases. The peak which will be analyzed in detail in panel B and C is highlighted in red. \textbf{B} Wavelength shift of the PL of a quantum dot (peak 5 of panel A) as a function of the phase between the PL pump and the infrared pump. The simulated wavelength shift is shown in red. The dotted lines show $\pm 3\sigma$ evaluated from the measurement of the same QD peak driven out of resonance. \textbf{C} Wavelength shift of the PL of the excitonic line labelled 5 in panel A, as a function of driving frequency at a fixed phase of 270$^\circ$. In red we show the simulated wavelength shift, considering the phase lag of the resonator when sweeping the driving frequency across the resonant frequency. The red shaded area in B and C indicates the uncertainty on the phase due to the pulse duration used for the stroboscopic measurement.}
    \label{fig:6}
\end{figure}

By fitting the excitonic lines of the peaks we observe two general behaviors. Some peaks (labeled 2, 3, 4) show a large drift over time, which is attributed to a thermal drift in the waveguide, where the QDs related to these lines are most probably located (see Supplementary for a more detailed discussion). The second family of peaks (1, 5) do not show this drift, indicating that they are positioned in a different nanobeam, which is not in thermal contact with the waveguide. Peak 1 does not show a statistically significant shift, suggesting that it originates from a QD in a low-strain region.
The behaviour of peak 5 is shown in Fig. \ref{fig:6}B where we plot the wavelength shift of such peak as a function of the driving phase. Here we observe a clear $2\pi$-periodic change correlated with the harmonic mechanical motion of the nanobeam, which we attribute to strain coupling.
To show statistical significance, we included two dotted lines which represent $\pm 3\sigma$ evaluated from repeated measurements of the same QD peak driven out of resonance.

To confirm that the shift we are observing is given by the strain modulation of the excitonic transitions of the QD, we fix the phase between the IR driving laser and the PL pump modulations to 270$^\circ$ and sweep the driving frequency across the mechanical resonance, as shown in Fig. \ref{fig:6}C. Here we clearly see a significant shift of the wavelength emission of the QD when driven close to resonance.

To properly analyze the data in Figs. \ref{fig:6}B and C we need to consider the phase lag of the resonator, which results in a $\pi$ shift when driven across the resonance.
Solving the differential equation describing the motion of a harmonic oscillator driven with a force $F(\omega) = F_0 \sin(\omega t + \phi)$, we get the following expression for the displacement $x$ at a time $t$:
\begin{equation}\label{eq:harmonicOscillator}
    x(\omega, \phi) = \frac{a}{ \sqrt{(\omega_\mathrm{m}^2 - \omega^2)^2 + (2\zeta \omega \omega_\mathrm{m})^2}} \sin{(\omega t + \phi + \theta)}  
\end{equation}
with
\begin{equation} \label{eq:phase}
    \theta=\cos^{-1}\left[ \frac{\omega_\mathrm{m}^2 - \omega^2 }{\sqrt{(\omega_\mathrm{m}^2 - \omega^2)^2 + 4\zeta^2\omega^2\omega_\mathrm{m}^2 }} \right]
\end{equation}
in which $\zeta = 1/(2Q)$, $\phi$ is the general offset between the phase of the drive frequency of the resonator and the optical pulse, $a = F/m_\mathrm{eff}$, and $m_\mathrm{eff}$ is the effective mass of the resonator. The values of $\zeta$ and $\omega_\mathrm{m}$ are taken from the fit of the thermal peak, and correspond to $1.06\times 10^{-4}$ and $20.85$~MHz, respectively.
Since the energy shift of the QD is directly proportional to the displacement of the nanobeam, we can write the wavelength shift ($\Delta \lambda$) as:
\begin{equation}\label{eq:DeltaLambda}
    \Delta\lambda(\omega, \phi) = \frac{A}{ \sqrt{(\omega_\mathrm{m}^2 - \omega^2)^2 + (2\zeta \omega \omega_\mathrm{m})^2}} \sin{(\omega t + \phi + \theta)}
\end{equation}
where $A$ is a parameter that sets the amplitude of the modulation.

The predictions of this model are shown in Fig. \ref{fig:6}B and C as red lines. In panel B the driving frequency $\omega$ is set to the mechanical resonance $\omega_\mathrm{m}$. The value of $A$ has been chosen so that the amplitude of the modulation corresponds to 24~pm, and $\phi=235^\circ$. For these assumed parameter values, we see that the model can describe the observed variations in the experimental data.

To evaluate the model as a function of frequency and compare it to the data in panel C, we use the wavelength shift for different values of driving frequency (in the range from 20.82~MHz to 20.92~MHz) from Eq. \ref{eq:DeltaLambda}. In this case, we keep $\omega t$ constant, as the values in panel C correspond to a fixed phase delay between the driving laser and the PL pump (270$^\circ$). For the plotted model curve, all the other parameters ($A$, $\phi$, $\zeta$ and $\omega_\mathrm{m}$) are the same as those used in panel B.

The red shaded area in both panel B and C represents the uncertainty on the phase, which results from the time duration of the PL pump pulse.
The model shows generally a good agreement with the experimental data, including the oscillatory behavior as a function of phase in Fig. \ref{fig:6}B and the sharply resonant dependence on drive frequency in Fig. \ref{fig:6}C. We speculate that the observed discrepancies could be due to small drifts between measurements of the lensed fiber used to couple light into the device. This could result in fluctuations of the amplitude modulation, and also in small variations of the mechanical frequency.

\section{Discussion}
Integrating QDs into an optomechanical resonator gives rise to a tripartite system where coupling between the optical field and the exciton is mediated by phonons. The strength of such interaction for our device was experimentally measured by optically driving the mechanical resonator and measuring the shift in the excitonic lines of the QD. In this section we evaluate the strength of the strain coupling per single phonon and then discuss the regimes in which the coherent phonon-exciton coupling can be achieved.

\subsection{Vacuum strain coupling rate}
In Fig. \ref{fig:6}B we report the wavelength shift of the excitonic line of a QD modulated in resonance. The data shows a good agreement with a numerical model based on Eq. \ref{eq:DeltaLambda}, with an amplitude of the modulation corresponding to $4.9$~GHz, and a phase shift $\phi=235^\circ$ which can be related to the delay between the mechanical drive and the stroboscopic pump.
The mechanical drive is implemented by using an intensity-modulated laser tuned to the symmetric optical mode while a low power laser is used to monitor the mechanical vibration of the resonator. From the ratio between the area of the pump peak and the thermal peak (assuming a temperature of 10 K) we can then estimate the number of phonons in the resonant condition of \ref{fig:6}B, corresponding to $n_\mathrm{p} \approx 2.6\times 10^8$ (see Supplementary).
From this value we can derive the vacuum strain coupling rate \cite{Leijssen2017} as being $g_\mathrm{QD} ⁄ 2\pi = \Delta E/(h \sqrt{2 n_\mathrm{p}}) = 214$~kHz. This estimation matches with the expected range according to the simulation reported in Fig. \ref{fig:2}D. We note that the value we measure is in the same range as those reported by earlier works, with strain coupling spanning from 68 kHz to 3.6 MHz in different regimes and architectures \cite{Yeo2014, Montinaro2014, Munsch2017, Kettler2020, choquer_quantum_2022, Spinnler2024}.

\subsection{Perspectives}
When looking at the practical application of using the strain coupling as coherent phonon-exciton interface, the coherence of the interaction is of paramount importance. The metric to evaluate the coherence is the cooperativity $C=4g_\mathrm{QD}^2 / (\Gamma_\mathrm{m} \gamma_\mathrm{QD})$ with $\gamma_\mathrm{QD}/2\pi\sim$10~GHz being the total decoherence rate of the QD \cite{petruzzella2017tunable}. By driving the QD with a laser detuned by $\pm \omega_\mathrm{m}$ and operating in the resolved sideband regime, the interaction gets enhanced by the optical field. In this system it can be proven that the coupling rate between the phonons and exciton becomes $g_\mathrm{QD}^*= g_\mathrm{QD} \Omega_\mathrm{R} ⁄\omega_\mathrm{m}$ with $\Omega_\mathrm{R}$ being the laser Rabi frequency \cite{wilsonrae2004}. 
Such interaction can be implemented by using breathing modes (frequency typically in the range $\omega_\mathrm{m} / 2\pi \sim 3$~GHz, and losses $\Gamma_\mathrm{m}/2\pi\sim200$~kHz \cite{forsch_microwave--optics_2020}) as mechanical resonator. Numerical simulations of breathing modes (in Supplementary) confirm that $g_\mathrm{QD}/2\pi\sim4$~MHz is achievable for high-frequency modes.
Assuming $\Omega_\mathrm{R}/2\pi \approx 20$~GHz \cite{yoshie_vacuum_2004} the cooperativity of such a system reaches $C>1$. At sufficiently low environmental temperature, this could yield quantum coherent operation.

\section{Conclusions}
In this work we presented an approach to achieve strain coupling of a two-level system, specifically excitons in an InGaAs/GaAs QD, to an optomechanical cavity. The optomechanical cavity consists of two double-clamped nanobeams which are driven in resonance with an intensity modulated laser. Driving the device allows achieving a phonon population of $n_\mathrm{p}\sim10^8$ at cryogenic temperatures, enhancing the strain-induced energy shift of the excitonic emission of the QD. The coupling has been measured by performing stroboscopic photoluminescence measurements of the QD, and showing the coherent modulation of its resonance frequency with respect to the driving laser.
The use of optical drive allows implementing and testing a tripartite system which couples together photons, phonons and excitons.

We experimentally inferred a QD frequency shift of $4.9$ GHz, which corresponds to a vacuum strain coupling rate of $214$ kHz. Even though with our experimental conditions we cannot achieve quantum cooperativity greater than 1, with adjustments to the design to shift the mechanical frequency to the GHz domain and by driving the QD in the sideband, a quantum cooperativity greater than 1 could be within reach. This provides an intriguing prospect towards the use of phonons to coherently couple different quantum systems together.

\section*{Funding}
This research was funded by Netherlands Organization for Scientific Research (NWO/OCW), as part of the Vrij Programma (grant n. 680-92-18-04) grant and of the Zwaartekracht Research Center for Integrated Nanophotonics (grant no. 024.002.033)

\section*{Acknowledgments}
The authors thank Pierre Busi (TU/e) for his contribution to the design of PhCCs, J. F. A. Vonk (TU/e) for his contribution to the experimental methods and dr. Francesco Pagliano (nanoPHAB b.v.) for developing the fabrication recipe used in this work. The authors acknowledge useful discussions with J. del Pino (AMOLF), R. Burgwal (AMOLF) and the other members of the \textit{QUAKE} consortium.

\section*{Disclosures}
The authors declare no conflict of interest.

\section*{Data Availability Statement}
Data underlying the results presented in this paper may be obtained from the authors upon reasonable request.

\bibliography{bibliography}

%Reset figure counter, and change figure command to label figures with 'S<fignumber>' for supplementary.
\setcounter{figure}{0}
\setcounter{equation}{0}
\makeatletter 
\renewcommand{\thefigure}{S\@arabic\c@figure} 
\renewcommand{\theHfigure}{S.\thefigure} %ensures hyperref is also reset 
\renewcommand{\theHequation}{S.\theequation}
\makeatother
\onecolumngrid
\clearpage

\section*{Supplemental document}

\section{Driving a mechanical resonator through a modulated laser field}
When an optomechanical cavity is optically driven with a laser field, the radiation pressure force per single photon is given by \cite{aspelmeyer_cavity_2014}:
\begin{equation}
    F = \hbar G
\end{equation}
with $G = g_0/x_\mathrm{zpf}$.
The amplitude of the motion of the resonator $x$ due to an oscillating force $F(\omega)$ is $x(\omega) = \chi(\omega) F(\omega)$ with
\begin{equation}
    \chi(\omega) = \frac{1}{m_\mathrm{eff}(\omega_\mathrm{m} ^2 - \omega^2) - im_\mathrm{eff}\Gamma_\mathrm{m} \omega}
\end{equation}
which for resonant driving $(\omega = \omega_\mathrm{m})$ becomes:
\begin{equation}
    \chi(\omega_\mathrm{m}) = \frac{i}{m_\mathrm{eff}\Gamma_\mathrm{m} \omega_\mathrm{m}}
\end{equation}

The amplitude of the motion, when driven in resonance then becomes:
\begin{equation}
    |x(\omega_\mathrm{m})| = \frac{\hbar G n_\mathrm{cav}}{m_\mathrm{eff}\Gamma_\mathrm{m} \omega_\mathrm{m}} = \frac{2 g_0 n_\mathrm{cav}}{\Gamma_\mathrm{m}} x_\mathrm{zpf}
\end{equation}
where we make use of $x_\mathrm{zpf}=\sqrt{\hbar / (2 m_\mathrm{eff} \omega_\mathrm{m})} $ and assume that the laser intensity that gives rise to cavity occupancy $n_\mathrm{cav}$ is fully modulated.

The number of phonons can then be evaluated from the amplitude of the motion with:
\begin{equation}\label{eq: number of phonons}
    n_\mathrm{p} = \frac{|x|^2}{2 x_\mathrm{zpf}^2} = \frac{2 g_0^2 n_\mathrm{cav}^2}{\Gamma_\mathrm{m}^2}.
\end{equation}

\section{Thermal drift}
Given the random distribution of the QDs and their low surface density it is expected that most of the excitonic lines would not show signs of strain coupling. Indeed, more than 20 devices were measured, and for each of them 2-3 spots were measured, but only on two devices a QD that shows signs of strain coupling was identified. One is not reported in this manuscript as the observed synchronous shift was small compared to the measurement uncertainty.

In the device described in the main text, besides the excitonic line (identified as peak 5) which shows strain coupling, other 4 peaks are visible, 3 of which clearly blueshift over time (measurements were taken starting from phase 0$^\circ$ to phase 360$^\circ$). To study this unexpected behavior, we plot the wavelength shift of such peaks over the time employed to carry out the entire study (Fig. \ref{fig:1}).
\begin{figure}[h]
    \centering
    \includegraphics[width=.7\linewidth]{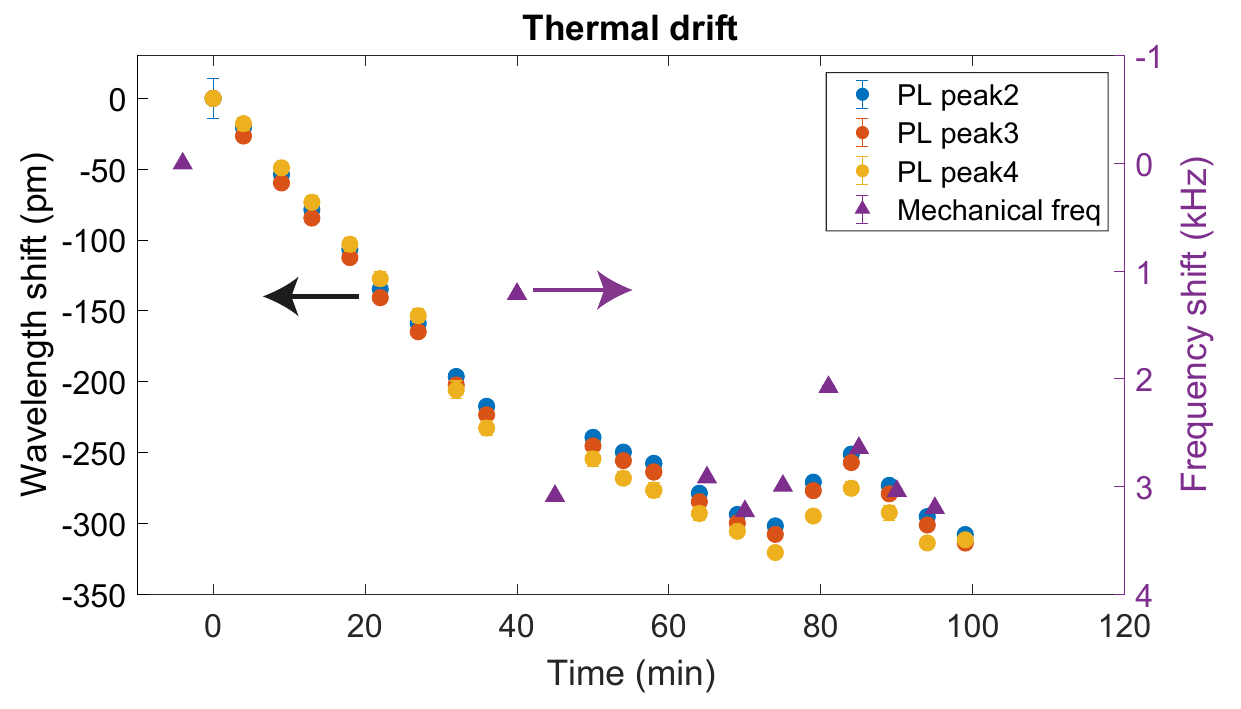}
    \caption{Thermal drift over time. Left axis refers to the wavelength shift measured in the PL of the QDs (circles), while the right axis shows the shift of the mechanical resonance (triangles).}
    \label{fig:S1}
\end{figure}

In the same plot we also include the datapoints related to the mechanical resonance shift, to check whether the observed drift can be related to thermal effects.
From the plot we see that both the wavelength shift and the mechanical frequency are compatible with a slow decrease of the local temperature over 100 minutes. Particularly, from a blueshift of the QDs of $\sim300$~pm we can estimate a change in temperature of 10 K \cite{Faraon2007}, while a mechanical shift of $\sim3$~kHz corresponds to a mean temperature change of 3 K \cite{Burenkov1973} of the entire resonator. The mismatch between the two values suggests that the thermal shift is local and caused by a drift of the optical setup which slightly moves the pump laser across the sample, locally changing the temperature of the QDs, while the temperature of rest of the device remains stable over time. We can then conclude that the excitonic lines 2, 3 and 4 are related to QDs located in the waveguide, thermally isolated from the cavity, while the excitonic lines 1 and 5 are related to QDs located in the nanobeams.

\section{Estimation of the driven phonon number} \label{sec: estimation np}
By performing a pump-probe measurement the amplitude of the pump peak can be measured and compared with the thermal peak (Fig. \ref{fig:2}). The pumping conditions are kept the same as the strain-coupling experiment.
There the data can be split into two parts, so that the pump peak (Gaussian) and the thermal background (Lorentzian) can be fitted independently. The area of the thermal peak $A_\mathrm{th}$ is proportional to the number of thermal phonons $n_\mathrm{th}=(e^{\hbar\omega_\mathrm{m} ⁄ k_B T} - 1)^{-1}$ while the area of the pump peak $A_\mathrm{pump}$ is proportional to the number of pumped phonons $n_\mathrm{p}$. The latter can then be simply evaluated as:

\begin{equation}\label{eq: np}
    n_\mathrm{pump} = \frac{A_\mathrm{pump}}{A_\mathrm{th}}n_\mathrm{th}
\end{equation}

\begin{figure}[h]
\centering\includegraphics[width=.7\linewidth]{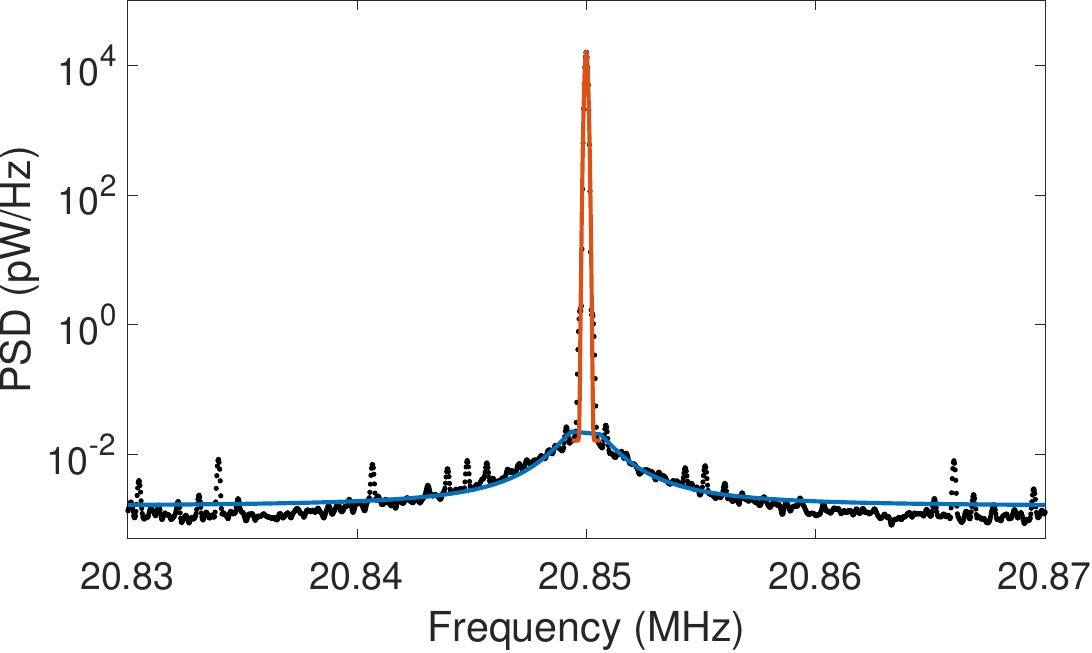}
\caption{Pump-probe measurement showing the driven mode on top of the thermally excited resonance.}
\label{fig:S2}
\end{figure}

To evaluate the number of thermal phonons in the mechanical resonator, we need to make an assumption on the temperature of the cavity. As the cold stage is at around 8 K, we assume that the temperature of the device is around $\approx$ 10 K, including possible heating effects due to the IR pump laser. Under this assumption, the thermal population for the differential mechanical mode is $n_\mathrm{th}\approx 10^4$, and by using the data obtained from the fit, we get $n_\mathrm{p}\approx 2.6\times10^8$.

\section{Estimation of the vacuum optomechanical coupling}
The value of the vacuum optomechanical coupling can be estimated from Eq. \ref{eq: number of phonons} by driving the cavity in resonance and measuring the number of driven phonons independently, as shown in Sec. \ref{sec: estimation np}.
By combining Eq. \ref{eq: number of phonons} and Eq. \ref{eq: np} we can write:
\begin{equation}
    \frac{A_\mathrm{pump}}{A_\mathrm{th}} = \left(\frac{2g_0^2}{\Gamma_\mathrm{m}^2 n_\mathrm{th}}\right)n_\mathrm{cav}^2
\end{equation}
which shows a quadratic dependence on the number of photons in the cavity.
The latter can be estimated measuring the power in the lensed fiber, knowing the extrinsic losses of the cavity $\kappa_\mathrm{ex} \approx \kappa/2 = 2\pi \times 73.1$~GHz and the coupling efficiency between the lensed fiber and the waveguide ($9.6\%$). To avoid thermal shifts between the measurements, the average power in the cavity was kept constant (the laser was set to 7.5~$\mu$W) while the amplitude of the modulation was swept from 1.5 to 545~$\mu$W.

The measurements and their fit are presented in Fig. \ref{fig:4}.
\begin{figure}[h]
    \centering
    \includegraphics[width=.7\linewidth]{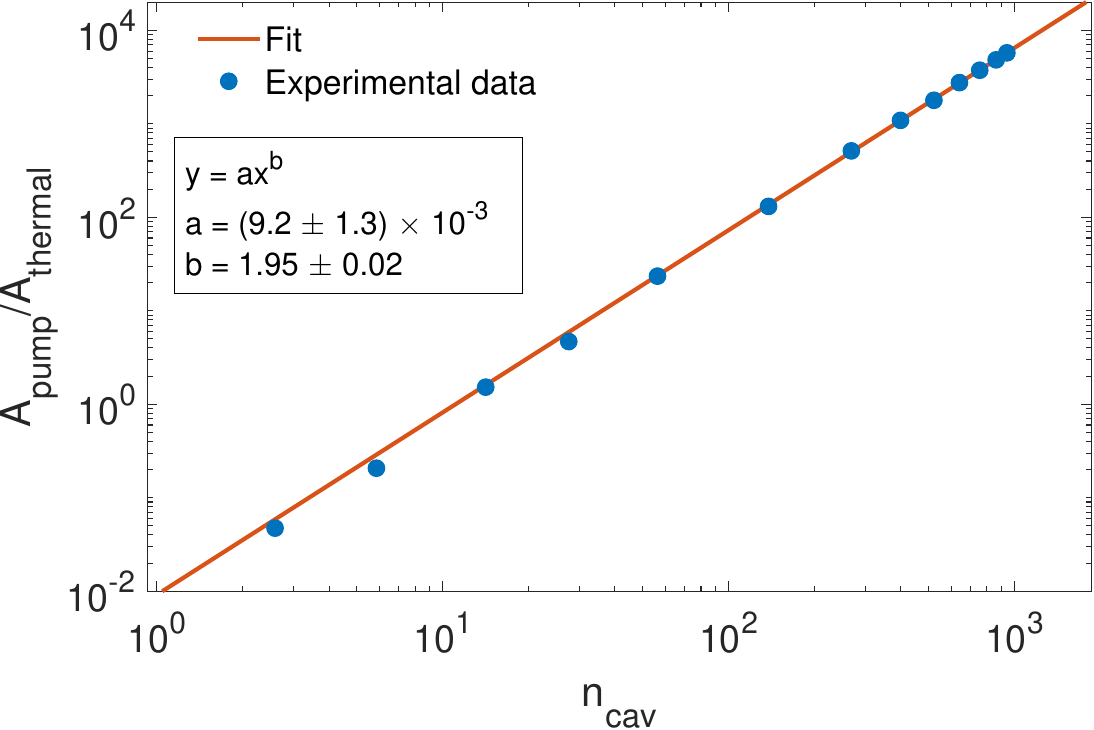}
    \caption{Rate of the amplitude of the driven mechanical peak and the thermal peak as a function of the number of photons in the cavity. The equation of the fit and the fitting parameters are shown in the inset.}
    \label{fig:S4}
\end{figure}
This experiment was carried out at room temperature, so $n_\mathrm{th}=3.1\times 10^5$. 
The mechanical loss rate can be evaluated from a Lorentzian fit of the mechanical spectrum, and corresponds to $\Gamma_\mathrm{m}/2\pi = 9.6$~kHz.
From the fit we can finally extract the value of $g_0/2\pi = 363\pm25$~kHz. 
A numerical simulation of the cavity gives a vacuum optomechanical coupling corresponding to 427~kHz. We attribute the discrepancy between the experimental value and the simulated value to differences in the geometry of the fabricated structure with respect to the design.

\section{Simulation of vacuum strain coupling in a high-frequency mechanical resonator}
In Sec. 4.2 in the main text, we propose a different approach which could lead to a cooperativity greater than 1.
In Fig. \ref{fig:5} we show an optomechanical structure which could be used to increase the strain coupling rate.
\begin{figure}
    \centering
    \includegraphics[width=0.9\linewidth]{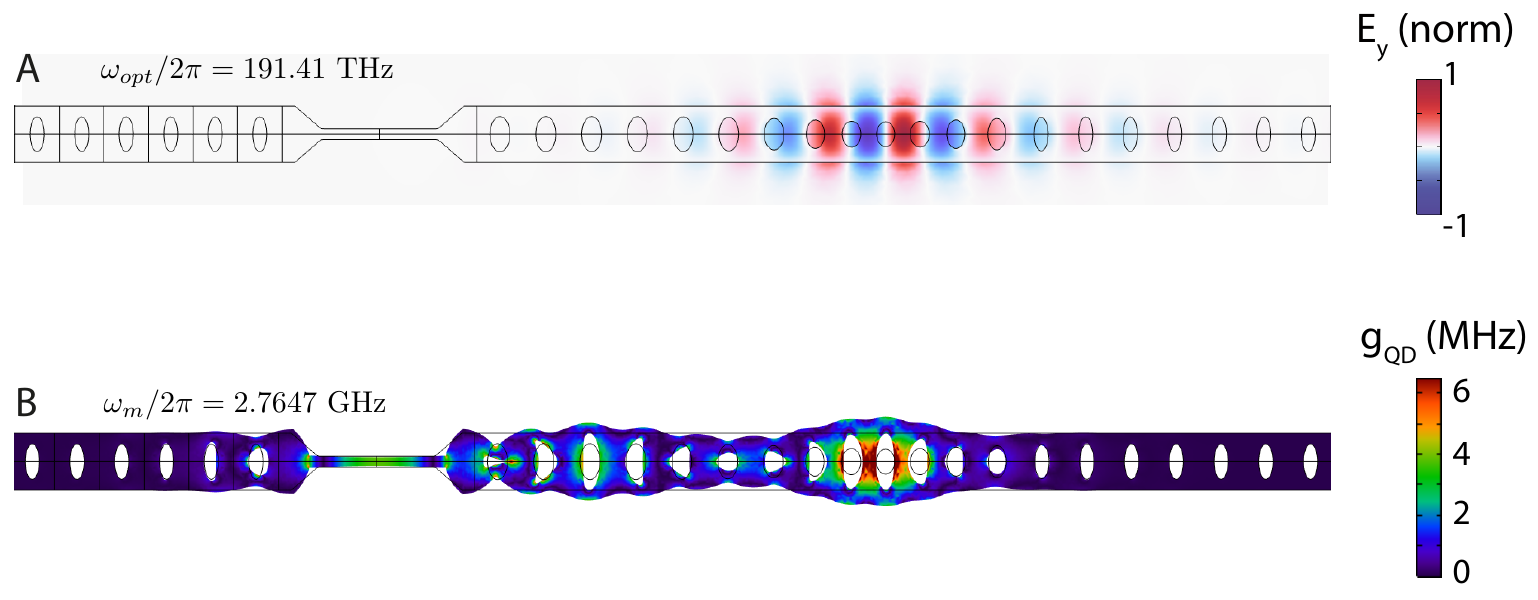}
    \caption{\textbf{A} Simulation of the fundamental optical mode with frequency $\omega_\mathrm{opt}/2\pi = 191.41$~THz and $Q_\mathrm{opt}=4.8\times10^4$. \textbf{B} Simulation of the  mechanical mode with resonant frequency 2.7647~GHz and $Q_\mathrm{m} = 1.9\times10^6$. The color scale shows the profile of the vacuum strain coupling, corresponding to a value of about 4~MHz in the center of bowtie for both modes.}
    \label{fig:S5}
\end{figure}
It consists of an optomechanical cavity coupled to a bowtie resonator. The mechanical coupling is achieved via a phononic waveguide, i.e. a photonic/phononic crystal which has an optical bandgap (to confine the optical mode in the optomechanical cavity) but does not have a mechanical bandgap at the mechanical frequency of interest.
In this device, the optomechanical cavity can be used to drive the mechanical mode by means of a modulated laser light. The QD energy shift can then be measured in the bowtie region, which is engineered to confine and enhance the strain in its center.

We provide both an optical and a mechanical simulation, in which the vacuum strain coupling profile is shown to reach values in the order of $\sim3$~MHz in the center of the bowtie. This structure is significantly more challenging to fabricate, due to the critical frequency matching between the two mechanical resonators. For this reason, the simpler structure described in the main text was used in this work.

\section{Nanobeam parameters} 
The parameters of the photonic crystals employed in the work are reported in Tab. \ref{tab:nanobeam} and Tab. \ref{tab:waveguide}.
Each hole of the PhCC and the waveguide is identified with a number, use Fig. \ref{fig:6} as a reference.
\begin{table}
    \centering
    \begin{tabular}{|cccc|}
        \hline
        \# hole & Lattice parameter (nm) &  Minor axis (nm) &	Major axis (nm) \\
        \hline
        1 & 374.375 & 106.875 & 360.156 \\
        2 & 374.375 & 106.875 & 360.156 \\
        3 & 373.750 & 106.875 & 359.375 \\
        4 & 373.125 & 108.125 & 358.594 \\
        5 & 370.625 & 110.000 & 354.688 \\
        6 & 365.625 & 115.000 & 349.219 \\
        7 & 356.875 & 123.125 & 337.500 \\
        8 & 346.250 & 133.750 & 321.094 \\
        9 & 335.625 & 146.250 & 301.563 \\
        10 & 328.750 & 157.500 & 282.813 \\
        11 & 326.875 & 161.875 & 279.688 \\
        12 & 328.750 & 157.500 & 282.813 \\
        13 & 335.625 & 146.250 & 301.563 \\
        14 & 346.250 & 133.750 & 321.094 \\
        15 & 356.875 & 123.125 & 337.500 \\
        16 & 365.625 & 115.000 & 349.219 \\
        17 & 370.625 & 110.000 & 354.688 \\
        18 & 373.125 & 108.125 & 358.594 \\
        19 & 373.750 & 106.875 & 359.375 \\
        20 & 374.375 & 106.875 & 360.156 \\
        21 & 374.375 & 106.875 & 360.156 \\
        \hline
    \end{tabular}
    \caption{Parameters of the photonic crystal holes for the structure investigated in the main text.}
    \label{tab:nanobeam}
\end{table}

\begin{table}
    \centering
    \begin{tabular}{|cccc|}
        \hline
        \# hole & Lattice parameter (nm) &  Minor axis (nm) &	Major axis (nm) \\
        \hline
        1  & 374.375 & 74.219  & 89.790  \\
        2  & 374.375 & 78.177  & 123.710 \\
        3  & 374.375 & 82.135  & 157.631 \\
        4  & 374.375 & 86.094  & 191.551 \\
        5  & 374.375 & 90.052  & 225.472 \\
        6  & 374.375 & 94.010  & 259.392 \\
        7  & 374.375 & 98.958  & 293.313 \\
        8  & 374.375 & 102.917 & 326.235 \\
        9  & 374.375 & 106.875 & 360.156 \\
        10 & 374.375 & 106.875 & 360.156 \\
        11 & 374.375 & 106.875 & 360.156 \\
        12 & 374.375 & 106.875 & 360.156 \\
        13 & 374.375 & 106.875 & 360.156 \\
        14 & 374.375 & 106.875 & 360.156 \\
        15 & 374.375 & 106.875 & 360.156 \\
        16 & 374.375 & 106.875 & 360.156 \\
        17 & 374.375 & 106.875 & 360.156 \\
        \hline
    \end{tabular}
    \caption{Parameters of the photonic crystal holes in the waveguide.}
    \label{tab:waveguide}
\end{table}

\begin{figure*}[h]
    \centering
    \includegraphics[width=1\linewidth]{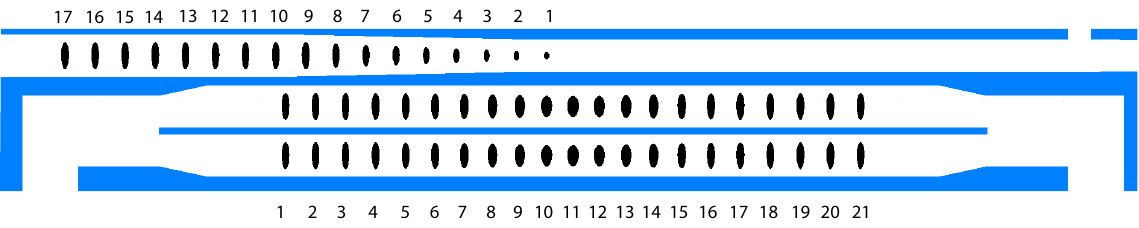}
    \caption{Enter Caption}
    \label{fig:6}
\end{figure*}
The other parameters of the design are shown in the sketch in Fig. \ref{fig:3}, all units are nm.
\begin{figure*}
    \centering
    \includegraphics[width=0.9\linewidth]{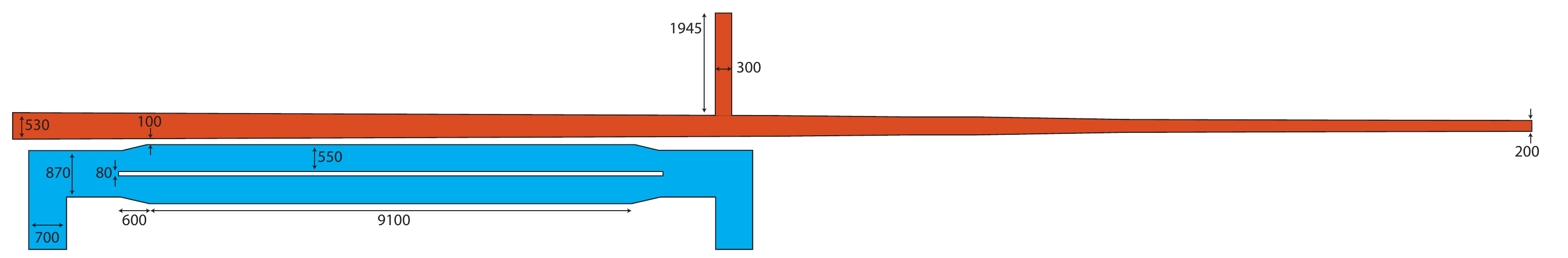}
    \caption{Sketch reporting all the parameters of the employed design. All values are expressed in nm.}
    \label{fig:S3}
\end{figure*}

% Bibliography

\end{document}